\documentclass{elsart}
\usepackage{graphicx}
\usepackage{latexsym}
\usepackage{amssymb}
\begin{document}
\begin{frontmatter}

\title{Profile driven interfaces in 1 + 1 dimensions : periodic steady
states, dynamical melting and detachment }

\author{Abhishek Chaudhuri and Surajit Sengupta}

\address{Satyendra Nath Bose National Centre for Basic Sciences,
Block-JD, Sector-III, Salt Lake,
Calcutta - 700098.}

\begin{abstract}
We study the steady state structure and dynamics of a 2-d Ising interface
placed in an inhomogeneous external field with a sigmoidal profile which moves 
with velocity $v_{e}$. In the strong coupling limit the problem maps onto an 
assymmetric exclusion process involving motion of particles 
in 1-d with position dependent right and left jump probabilities. For small 
$v_{e}$, the interface is stuck to the field profile. As $v_{e}$ increases 
the profile detaches from the interface. At the transition
point( and beyond ), the interfacial structure and dynamics is characterized 
by KPZ exponents. For small $v_{e}$, on the other hand, the interface is 
macroscopically 
smooth with a vanishing roughness exponent $\alpha$. The interfacial structure 
is periodic with a periodicity which depends on the orientation of the 
interface. For a fixed orientation this periodic structure ``melts'' as $v_e$ 
is increased. We determine the dynamical ``phase - diagram'' of this system 
in the $v_e$ - orientation plane.  
\end{abstract}
\begin{keyword}
Interface Dynamics, Kinetic Ising model, Dynamical phase transition
\PACS 05.10Gg,64.60.Ht,68.35.Rh
\end{keyword}
\end{frontmatter}

\section{Introduction}

Consider a one-dimensional (1-d) Ising interface {\cite{Barabasi,Will}} 
between up and down 
spins in two dimensions (2-d) obeying single-spin flip Glauber 
dynamics {\cite{Men,Abraham}}. In the presence of a uniform driving 
field {\cite{musanan,Ferreira,M.Barma,Maj1,Maj2,Kardar}} the interface 
moves with a velocity which depends on the magnitude (and sign) 
of the driving field. On the other hand, a fixed external 
field profile which is positive in the region of up spins and negative in 
the region of down spins, would stabilize a stationary, macroscopically flat 
interface.  
In this paper we study systematically the structure and dynamics of 
this Ising interface as this field profile is moved with an arbitrary
velocity $v_{e}$.\\ 
There are two reasons why we are interested in this problem :
Firstly, there are several important practical examples 
where inhomogeneous fields drive interfaces. Some of them include 
zone purification of Si where the controlled motion of a temperature field 
profile is used to preferentially segregate impurities\cite{Hasen}, magnetization of a 
bar of iron with a permanent magnet, phase transitions induced by a travelling heat 
(welding) or pressure (metamorphosis of rocks) fronts etc. Secondly, we would
like to extend this study, in the future,  to the dynamics of solid interfaces 
where interfacial degrees of freedom are coupled to hydrodynamic 
modes of the bulk solid eg. phonons (acoustic emmissions\cite{Hasen}) and 
defects\cite{ourPRL}.
A systematic study of how these modes are excited in sequence as $v_{e}$
is increased is of great fundamental interest.\\
The object of our study here is the interface between up and down spin phases
(Fig. 1) in the limit $h/J , T/J \rightarrow 0$, where $J$ is the Ising 
exchange coupling, $T$ the temperature and the field $h(x,t)$ here 
is inhomogeneous, $h = {\it h_{max}}$ in the region where the magnetization 
is positive and $-{\it h_{max}}$ in region where it is negative separated by 
a relatively sharp {\it edge}. The {\it edge} of the field (i.e. where the 
field changes sign) lies at $S_e$. The {\it front} or interface, u(y,t) 
(no overhangs !) separates up and down spin phases. The interface 
is a bold curved line with the average position  $S_f$. 
To move the interface we move the edge with velocity ${\it v_e}$; 
in response the front moves with velocity ${\it v_f}$. Parts of the front 
which leads (lags) the edge of the 
field experience a backward (forward) force pulling it towards the edge.  
The driving force therefore varies in both space and time and depends on the 
relative position of the front compared to that of the edge of the dragging 
field. What is the behaviour of the front velocity $v_f$ as a function of $v_e$?
What is the structure of the interface in various regimes? These are the questions 
we address in this paper.

\begin{figure}[t]
\begin{center}
\includegraphics[width=6cm]{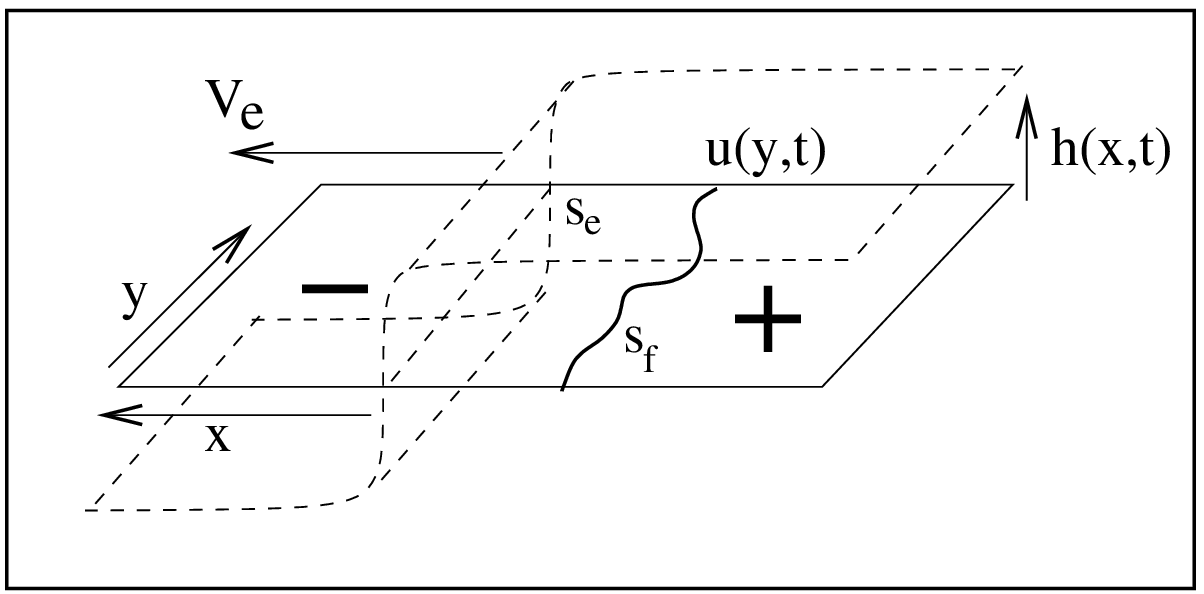}
\end{center}
\caption{
An Ising interface $u(y,t)$ (bold curved line) between regions of positive 
(marked $+$) and
negative (marked $-$) magnetization in an external, inhomogeneous field 
with a profile which is as shown(dashed line). The positions of the edge of the
field profile and that of the front are labelled $S_e$ and $S_f$ respectively.
}
\label{fig1}
\end{figure}

Briefly, our results are as follows
\begin{enumerate}
\item For any orientation of the interface, $v_f = v_e$ for small 
$v_e$ the front moves along with the field profile. We call this the 
``stuck'' phase.

\item At a velocity $v_{e} =  v_e^*$,
the front detaches from the field profile. At higher values of $v_e$ the front 
experiences an {\it uniform} magnetic field $h = h_{max}$ and the problem 
reduces to the growth of an Ising interface driven by a uniform field
\cite{musanan,Ferreira,M.Barma,Maj1,Maj2,Kardar}

\item The structure of the ``stuck'' interface is flat with a roughness exponent
$\alpha = 0$. Depending on the orientation of the interface, the height of the 
interface $u(y,t)$, as a function of $y$ and 
time $t$, may show periodic oscillations in $y$ and/or $t$. The nature of these 
oscillations depends crucially on the system size in a manner to be explained 
below.
\end{enumerate}
In the next section we introduce a continuum description of the problem and 
derive the relevant coarse grained equations of motion. In Section 3 we 
present the mean field solution to these equations. In Section 4 we introduce
fluctuations through an exact mapping to an assymmetric exclusion (particle
hopping) model in 1-d and present, analyze and discuss our results obtained 
from computer simulations of this model. In Section 5 we present our 
conclusions.

\section{\bf Continuum Description}

Let the magnetization of the 2-d Ising system be given by, 
$\phi \equiv \phi(x,y,t)$. We assume that for 
 $h/J , T/J \rightarrow 0$ the magnetization is uniform everywhere 
except near the interface which may be parametrized by a function 
$u(y,t)$, where $u$ is the height of the interface perpendicular to $y$. Hence 
the magnetization $\phi = \phi(x - u(y,t))$. The field profile is given
by $h = h_{max} \tanh ((x - v_{e}t)/\chi)$ where $\chi$ is the
width of the profile.
Model A dynamics\cite{Chaikin} for $\phi$ then implies,

\begin{eqnarray}
\frac{\partial \phi}{\partial t} &=& -\Gamma \frac{\delta H_{T}}{\delta \phi}
+ \zeta ({\bf r},t) 
\label{modelA}
\end{eqnarray}

where 

\begin{eqnarray}
H_{T} &=& \int d{\bf r} [ a_{1}\phi^{2} + a_{2}\phi^{4} + 
a_{3}(\nabla \phi)^{2} - h(x,t)\phi]
\end{eqnarray}

is the Hamiltonian and $\zeta$ is a Gaussian white noise with zero mean
and  

\begin{eqnarray}
<\zeta ({\bf r},t)\zeta ({\bf r^{\prime}},t^{\prime})> &=& 2k_{B}T{\Gamma}
{\delta({\bf r - \bf r^{\prime}})}{\delta(t - t^{\prime})}
\end{eqnarray}

Using $H_{T}$ in Eq.\ref{modelA}, taking $x - u(y,t) = v$ and converting all 
derivatives to derivatives over the profile $u(y,t)$, we have, 

\begin{eqnarray}
-\phi^{\prime}(v) \frac{\partial u}{\partial t} = &-&\Gamma [2a_{1}\phi(v)
 + 4a_{2}\phi^{3}(v) - 2a_{3}\phi^{\prime \prime}(v) + 2a_{3}\phi^{\prime}(v)
\frac {\partial^{2} u}{\partial y^{2}} 
\nonumber\\
&-& 2a_{3}\phi^{\prime \prime}(v)
(\frac{\partial u}{\partial t})^{2} - h(x,t)] + \zeta ({\bf r},t)
\end{eqnarray}

We then choose a $\phi$ dependent mobility $\Gamma$ contributing to the lowest
order in $\phi$ consistent with symmetry viz.
$\Gamma = \Gamma_{0} + \Gamma_{1}(\nabla \phi)^{2}$.
Substituting for $\Gamma$ and integrating both sides of the equation with 
respect to $x$ between limits $(u - \chi/2)$ and $(u + \chi/2)$ i.e.
over the interfacial region, remembering that $\phi$ has a sigmoidal 
profile, we finally get an equation of motion for the profile $u$. 

\begin{eqnarray}
\frac{\partial u}{\partial t} &=& \lambda_{1} \frac{\partial^{2} u}
{\partial y^{2}} - \lambda_{2} \Big(\frac{\partial u}{\partial y} \Big)^{2} 
\tanh \Big(\frac{u - v_{e}t}{\chi} \Big) - \lambda_{3}\tanh \Big(\frac{u - v_{e}
t}{\chi} \Big) + \zeta^{\prime} (u,t)
\label{m-KPZ}
\end{eqnarray}

where $\lambda_{1}$,$\lambda_{2}$ and $\lambda_{3}$ are constants. This 
is different from the familiar KPZ equation {\cite{Kardar}} in the fact
that it lacks 
Galilean invariance (
$u^{\prime} \rightarrow u + \epsilon y,\>\> y^{\prime} \rightarrow y - 
\lambda_{2}\epsilon t, \>\> t^{\prime} \rightarrow t$ ) 

In general crystal field effects introduce a lattice periodic force
\cite{Barabasi} which may be accounted for by including an additional term 
$V_{0}\sin(2 \pi u/a)$ ($a$ is the lattice parameter) to the right side 
of the above equation of motion.

\section{\bf Mean Field Result}

A mean field calculaton amounts to taking $u \equiv u(t)$ i.e. neglecting 
spatial fluctuations of the interface. Then 

\begin{eqnarray}
\frac{\partial u}{\partial t} &=& -\lambda_{3}\tanh \Big(\frac{u - v_{e}t}
{\chi} \Big) + \zeta^{\prime} (u,t)
\end{eqnarray}

For large times ($t \rightarrow \infty$), $u \rightarrow v_{f}t$, where 
$v_{f}$ is the average velocity of the front. Thus $v_{f}$ is obtained
by solving the self-consistency equation,

\begin{eqnarray}
v_{f} &=& -\lambda_{3}\tanh \Big(\frac{(v_{f} - v_{e})t}{\chi} \Big)
\label{mft-v}
\end{eqnarray}

In the $t \rightarrow \infty$ limit the hyperbolic tangent is replaced by the 
simply the sign of $v_{f} - v_{e}$ and is equivalent to taking 
$\chi \rightarrow 0$
namely, an infinitely sharp field profile. For small $v_e$ the only solution to 
the self-consistency equation is $v_f = v_e$ as can easily be verified 
graphically. For large edge velocities $v_{e} > v_{e}^*$, where 
$v_{e}^{*} = \lambda_{3}$ we get 
$v_{f} = \lambda_{3} = v_{e}^{*}$. We thus have a sharp transition 
from a region where the front follows the edge with the same velocity to 
one where it moves with a constant velocity unable to follow it anymore.
The interface velocity relaxes to its steady state value $v_{e}$ as $1/t$ 
in the region of low $v_{e}$.
The region where the front moves with a 
constant velocity is evidently the well studied problem of driving an 
interface by a homogeneous field \cite{Ferreira,Maj2}.\\ 
How is this result altered by including spatial fluctuations of $u$ ?
In order to answer this question we have mapped this interface model 
to an assymmetric exclusion process \cite{Ligget} and study
the dynamics both analytically and numerically using computer simulations.

\section{Beyond Mean Field Theory}
The mapping to the exclusion process follows \cite{Maj1} by considering 
$N_p$ particles distributed among $N_s$ sites of a 1-d lattice. 
The particles are labelled $n = 1,2,.......,N_p$
sequentially at $t = 0$. Any configuration of the system is specified 
by the set of integers 
$\{y(n)\}$ where $y(n)$ denotes the location of the $n$th particle. In the
interface picture $n$ needs to be interpreted as a horizontal
coordinate ($y$ in Fig. 1), and $y(n)$ as a local height $u(y,t)$. 
Each configuration $\{y(n)\}$ then defines a one-dimensional interface inclined
 to the horizontal with mean slope $1/\rho$ where $\rho = N_{p}/N_{s}$. The 
interface coordinates satisfy
$y(n+1) \geq y(n) + 1$, and periodic boundary conditions amount to setting
$y(n+N_p) = y(n) \pm N_s$. Motion of the interface under the influence
of a driving field corresponds to the hopping of particles. In each time step
($N_{p}$ attempted hops),
$y(n)$ tends to increase (or decrease) by 1 with probability $p$ (or $q$); it
actually increses (or decreases) if and only if ${y(n+1) - y(n) > 1}$. In our 
case the right and left jump probabilities $p$ and $q$ ($p+q = 1$) are not 
constants but themselves depend on the relative position of the interface 
$y(n)$ and the edge of the field $n/\rho + v_e t$. Note that in calculating 
this relative position we have to use the actual position of the interface 
{\em without}
periodic boundary conditions. In  our calculations reported here we 
use a bias $\Delta = p-q = \Delta_0{\rm sign}(y(n)- n/\rho - v_e t)$ with 
$\Delta_0 = 1$ unless otherwise stated. 
We are interested in the average vertical velocity of the interface 
$v_f$ defined as the total number of particles moving right per time 
step. In addition to the front velocity, we also examine the behaviour of 
the width of the interface: 
\begin{eqnarray}
\sigma^{2}(t) &=& N_p^{-1}\sum_{n=1,N_p}<(y(n,t) - y(n,0) - v_ft)^2>
\end{eqnarray}
as a function of time and system size $N_s$. 
The angular brackets denotes an average over the realizations of the 
random noise.\\
Note that the usual particle hole symmetry for an exclusion process 
\cite{musanan}
is violated since exchanging particle and holes changes the relative position
of the interface compared to the edge of the field.

\subsection{Monte Carlo simulations for the dynamical transition} 

We perform Monte Carlo simulations of the exclusion process using a random 
sequential update \cite{exclu} to understand the
behavior of the interface in the presence of the inhomogeneous field 
profile. We study the system for  different system 
sizes and  densities and obtain the average velocity $v_f$ of the steady 
state interface as a function of the velocity of the edge of 
the field profile $v_e$. The velocity $v_f$ is obtained by dividing the distance
moved by the interface for a certain (large) number of time steps by the total 
number of time steps after discarding the first few thousand steps to 
remove transients. Fig. 2 shows a sharp dynamical
transition from an initially ``stuck'' interface with $v_f = v_e$ to a free, 
detached interface with $v_f = v_e^{*} = \Delta_0 (1-\rho)$ the result 
for an assymetric exclusion process with density $\rho$. Note that 
the mean field solution for the front velocity and the dynamical transition 
is exact.
\begin{figure}[t]
\begin{center}
\includegraphics[width=6cm]{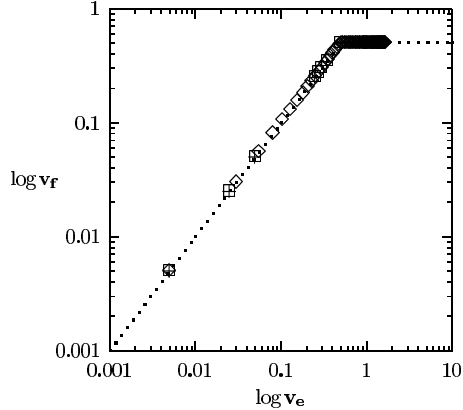}
\end{center}
\caption{The front velocity $v_f$ as a function of the velocity of the
dragging edge $v_e$ for $N_s = 100(\Box), 1000(\Diamond), 10000(+)$ 
and $\rho = 0.5$. All the data (symbols,$\Box,\Diamond,+$) collapse on the 
mean field solution (dashed line).
}
\label{fig2}
\end{figure}

\subsection{The {\it Stuck} phase ($v_{e} < v_{\infty}$)}

The stuck phase is characterized by $v_f = v_e$ and $\sigma$ bounded.
To obtain the ground state of the interface in the presence of a stationary 
($v_e = 0$)
field profile it is sufficient to minimize $\sum_n (y(n)- n/\rho)^2$ which 
demands $y(n)$ to
follow the edge $n/\rho$ as closely as possible subject to the constraint that 
$y(n)$ be an integer. This ground state structure is always periodic for 
$\rho < 1/2$. For $\rho \ge 1/2$ this periodicity is destroyed for infinetisimal 
$v_e$. For densities which are of the form $1/k$ where $k$ is 
an integer, the result is particularly simple viz. $y(n) = k\,n$. This corresponds
to a particle-hole system where the particles form a 1-d lattice with a lattice 
parameter of $k$. For an arbitrary density (orientation) the ground state is 
still periodic 
over short distances but has long-period (possibly incommensurate) modulations. 
We verify this by calculating the 
pair distribution function $g(l) = (1/N_p (N_p-1)) \sum_{n,m} \delta_{l,(y(m)-y(n))}$,
where $n,m$ are particle indices and $m > n$. 
The Fourier transform of $g(l)$ shows prominent delta function peaks.\\ 
The dynamics of the interface for $v_e > 0$ depends on whether or not the 
system size is compatible with the lattice parameter $k$.  
If the system size $N_s$ is an exact multiple of $k$ then the particles which are 
separated by intervening regions of holes can move independently of each other in 
response to the local bias $\Delta$.  
Let $P_i$ be the probability ($\sum_{i=-\infty}^{\infty} P_i = 1$) of obtaining 
a particle (any particle) in state $i$ where $i=0$ corresponds to a particle 
which has not moved from its initial position and $i = s$ ($= -s$) corresponds 
to one which has moved $s$ integral lattice spacings to the right (left) 
of its original position. Fluctuations of the interface about the ground state 
correspond to these independent particle motions which cost energy if $v_e t$ 
is integral. Consider now that $i < v_e t < i+1$ the form of the bias
$\Delta$ implies that the probabilities $P_i$ satisfy the following set of 
master equations,
\begin{eqnarray}
\dot P_j & = & -P_j + P_{j+1} \,\,\,\,\,\,\,\,\,\,\,\,\,\,\,\,\,\,\,\,\, {\rm for\,\,\, j\, >\, i+1} \\\nonumber
\dot P_j & = & P_{j-1} - P_j + P_{j+1}\,\,\,\,\,\,\,{\rm for\,\,\, j\, =\, i,i+1} \\\nonumber
\dot P_j & = & -P_j + P_{j-1}\,\,\,\,\,\,\,\,\,\,\,\,\,\,\,\,\,\,\,\,\, {\rm for\,\,\, j\, <\, i}.
\end{eqnarray}

Noting that the average position $S(t) =\, <N_p^{-1}\sum_n(y(n,t)-y(n,0)>$ 
of the interface 
is given simply by $S(t) = \sum_{i=-\infty}^{\infty} i\,P_{i}(t)$ and 
$\sigma^{2}(t) = \sum_{i=-\infty}^{\infty} (i^2 -i)\,P_{i}(t)$ we obtain the 
results shown in Fig. 3(a) and 3(b). It is clear that these results match the 
corresponding ones obtained from Monte Carlo simulations exactly. The interface 
therefore 
follows the profile in a jerky fashion and the width of this interface oscillates
between fixed bounds. Thus although the structure of the moving interface 
corresponds more or less with the ground state periodic structure for small 
$v_e$, the velocity is oscillatory. 
\begin{figure}[t]
\begin{center}
\includegraphics[width=10cm]{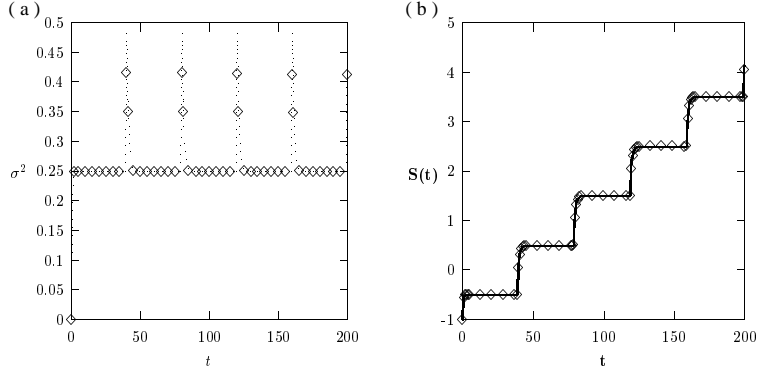}
\end{center}
\caption{(a)Variation of $\sigma^{2}$ with $t$ for for $\rho = 0.5$,
$v_{e} = 0.025$ and $p = 1.0$ and (b) variation of $S(t)$ with $t$ for 
$\rho = 0.5$, $v_{e} = 0.025$ and $p = 1.0$. Lines denote analytical
results while points denote monte carlo data.
}
\label{fig3}
\end{figure}
If however $\rho $ is not the reciprocal of an integer and the system size
does not accomodate an integral number of spatial periods of the ground 
state then the corresponding 1-d lattice contains long wavelength modulations
and the particle hoppings are not independent anymore. 
For fixed edge velocities
it is found that $\sigma^{2}$ is a constant in time independent of 
the size of the system, the constant, however, depends on 
$v_{e}$ and $\rho$. The average position of the interface does not show any 
oscillations and faithfully follows the field profile corresponding exactly
to the mean-field solution.

{\bf The Melting Transition}

The Fourier transform of $g(n)$, i.e. the structure factor $\tilde g(q)$, 
indicates a 
``melting transition'' of the periodic steady states with increasing  $v_{e}$
for any density ($\rho < 1/2$). We track this by plotting the  
intensity of the largest peak (smallest $q$) of $\tilde g(q_{max})$ 
as a function of $v_e$ for a number of densities. This is shown in 
Fig.4(a) . Also, the lattice paramater ($a$) obtained from $2{\pi}/q_{max}$
is used to determine the Lindemann ratio $L = {\sigma^{2}/a^{2}}$. 
The increase of the Lindemann ratio with the $v_{e}$ for different
densities (Fig. 4(b)) is another proof of a melting transition.
These results are summarized in the dynamical phase diagram (Fig. 5) 
for the Ising interface in a moving field profile. It shows an 
detachment transition along the line $v_e = (1-\rho)$ (for $\Delta_0 = 1$) 
and a melting transition. The exact position of this dynamical melting 
transition (unlike a thermodynamic transition) depends on the parameter used 
to characterize it. If we use $\tilde g(q_{max})$ then the melting 
transition occurs simultaneously with detachment for $\rho < 0.5$
and at $v_e = 0^+$ for larger $\rho$. Using the Lindemann parameter, however, 
one obtains a melting transition which preempts detachment.

\begin{figure}[t]
\hskip 3in
\begin{center}
\includegraphics[width=12cm]{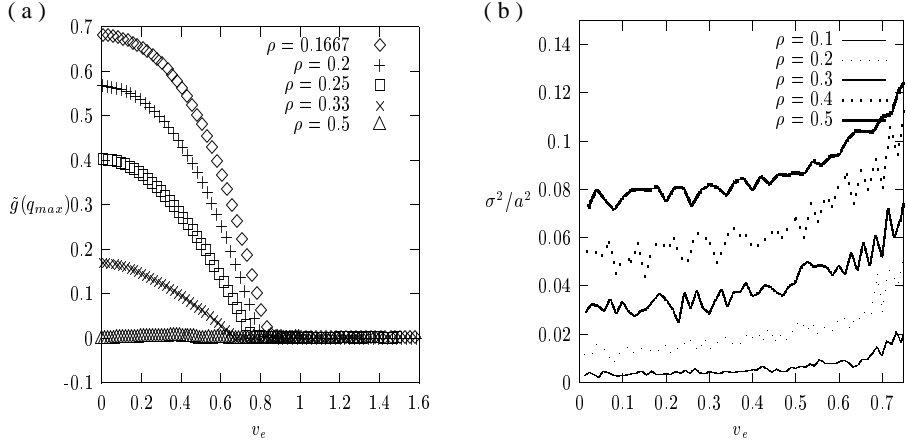}
\end{center}
\caption{(a) The structure factor $\tilde g(q_{max})$ vs. $v_e$ for various
$\rho$. Note that $\tilde g(q_{max})$ vanishes
as $v_e$ increases thereby implying a melting transition in the 1-d assymetric
exclusion process. (b) Variation of 
lindemann ratio $L$ with edge velocity $v_e$ for various densities. An increase
of $L$ again signifies melting. 
}
\label{fig4}
\end{figure}

\begin{figure}[t]
\begin{center}
\includegraphics[width=6cm]{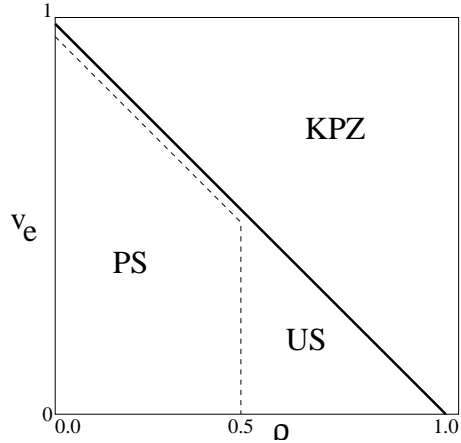}
\end{center}
\caption{Phase diagram for Ising interface driven by an inhomogeneous magnetic
field showing periodic stuck (PS) steady states for $v_e < (1-\rho)$ 
and $\rho < .5$, uniform stuck (US) steady states for $v_e < (1-\rho)$ 
and $\rho > .5$ and detached (KPZ) steady states for $v_e \ge (1-\rho)$.
}
\label{fig5}
\end{figure}

\section{\bf Behavior at the transition point}

We want to determine scaling form for $\sigma(t)$ at the transition 
point viz. the growth exponent $\beta$, the 
roughness exponent $\alpha$ and the dynamic exponent $z$. In the detached phase 
we know from renormalization group analysis that the 
exponents are in the Kardar-Parisi-Zhang (KPZ) universality class
\cite{Barabasi,Kardar} viz.
 $\beta = 1/3$, $\alpha = 1/2$  and $z = \alpha/\beta 
= 3/2$. To determine these exponents at the transition point we make use 
of {\sl Family-Vicsek scaling relation} \cite{Barabasi}

\begin{eqnarray}
\sigma(L,t) \sim N_s^{\alpha}f(t/N_s^{z})
\end{eqnarray}

Fig. 6 shows the variation of $t/N_s^{z}$ with $\sigma(L,t)/N_s^{\alpha}$ for 
different $p$ and different system sizes $N_s$. The curves collapse onto 
one curve once an intrinsic width $\sigma_{i}$, arising from finite-size and 
crossover effects \cite{Barabasi}, is substracted out. 
The exponents were found to be KPZ.
\begin{figure}[t]
\begin{center}
\includegraphics[width=10cm]{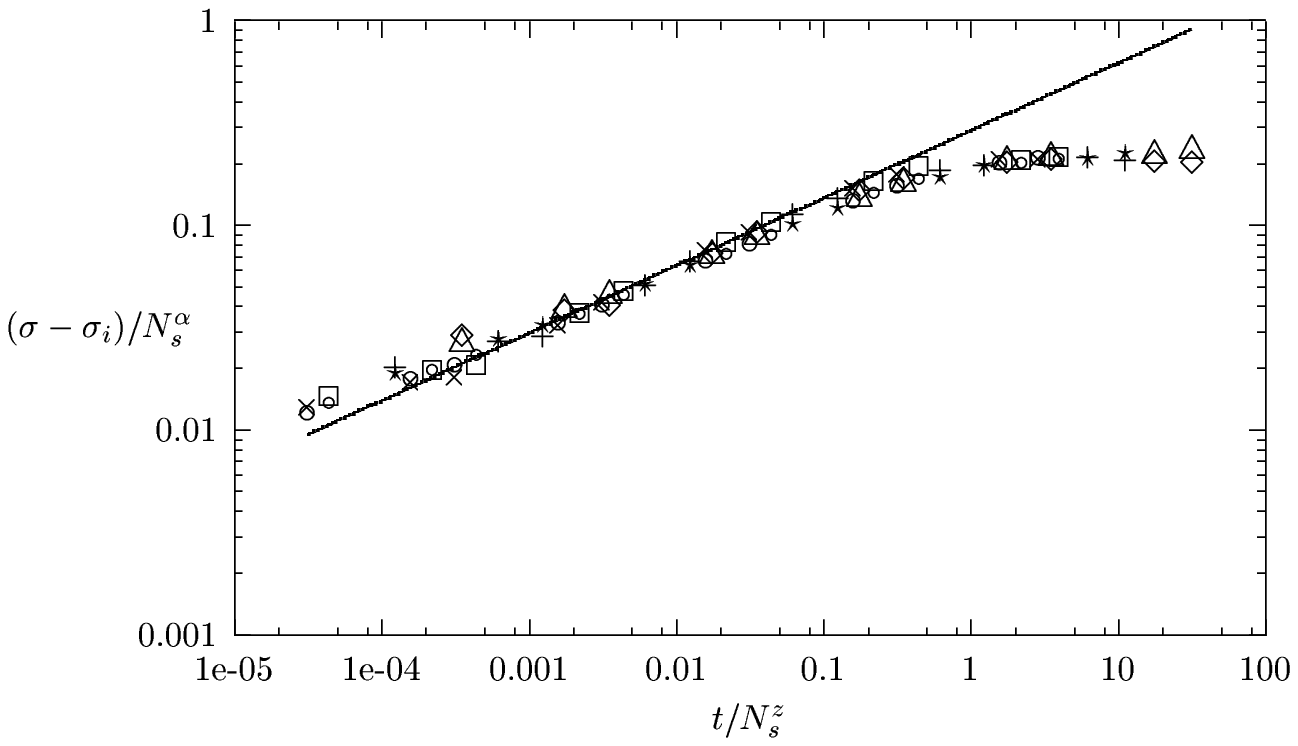}
\end{center}
\caption{Monte Carlo data $(\sigma-\sigma_i)/N_s^{\alpha}$ vs $t/N_s^{z}$ 
for $p = 1.0$, $p = 0.7$ and with $N_s = 1000$ [{$\times,\bigcirc$}] ,
$800$ [{$\Box,\circ$}] , $400$ [{$+,\star$}] , $200$ [${\Diamond,\triangle}$]. 
All the curves collapse to a single universal function showing KPZ scaling.
}
\label{fig6}
\end{figure}

To understand why this happens we go back to our modified KPZ equation 
(Eq.\ref{m-KPZ}) and make the transformation $u = u^{\prime} + v_{f}t$. We get

\begin{eqnarray}
\frac{\partial u^{\prime}}{\partial t} + v_{f} &=& \lambda_{1} 
\frac{\partial^{2} u^{\prime}}{\partial y^{2}} - 
\lambda_{2}\Big(\frac{\partial u^{\prime}}
{\partial y}\Big)^{2} \tanh\Big(\frac{(v_{f} - v_{e})t + u^{\prime}}{\chi}\Big) 
\nonumber\\
&&- \lambda_{3}\tanh \Big(\frac{(v_{f} - v_{e})t + u^{\prime}}{\chi}\Big)
+ \zeta^{\prime}(u,t)
\end{eqnarray}

Now substituting the mean field result for $v_f$ (Eq.\ref{mft-v}), making use 
of the fact that at the transition point $v_{f} = \lambda_{3}$ and simplifying
one can show that the above equation reduces to the familiar KPZ equation in 
$u^{\prime}$.

\section{Conclusion}
In this paper we have studied the static and dynamical properties of an 
Ising interface in 2-d subject to a non-uniform, time-dependent external 
magnetic field. The system has a rich dynamical phase diagram with 
several dynamical phases (steady states). The nature of these steady 
states depend on the orientation of the interface and the velocity 
of the external field profile. The detailed dynamics of the interface
depends on whether or not the size of the system is commensurate with 
the orientation. For a commensurate system, the interface follows the 
field in a jerky fashion and the width of the interface fluctuates 
between well defined bounds. For a general, incommensurate interface 
the motion of the interface is steady and the width is constant. For 
large velocities of the external field, the interface detaches from 
the profile and coarsens over time with KPZ exponents\\ 
In future we would like to study in detail 
further dynamical aspects of this system e.g. the hysteretic response of 
this system under time varying external parameters ($v_e$).

\noindent
{\bf Acknowledgement: }The authors would like to thank M. Barma, J. Krug, 
J.K. Bhattacharya, P. A. Sreeram and S. Majumdar for useful discussions.
One of the authors (A.C.) thanks the Council of Scientific and Industrial 
Research (C.S.I.R.), Government of India for a Junior Research Fellowship.


\end{document}